\documentclass[runningheads]{llncs}

\usepackage[T1]{fontenc}
\usepackage{graphicx}
\begin{document}

\title{Predicting At-Risk Programming Students in Small Imbalanced Datasets using Synthetic Data}
\titlerunning{Predicting At-Risk  Students using Synthetic Data}
\author{Daniel Flood\inst{1}\orcidID{0000-0002-2834-3473} \and
Matthew England\inst{1}\orcidID{0000-0001-5729-3420} \and
Beate Grawemeyer\inst{1}\orcidID{0000-0001-5583-8888}
}
\authorrunning{D. Flood, M. England and B. Grawemeyer}
\institute{Coventry University, United Kingdom \\
\email{\{Daniel.Flood, Matthew.England, Beate.Grawemeyer\}@coventry.ac.uk}}
\maketitle  % typeset the header of the contribution
\begin{abstract}
This study is part of a larger project focused on measuring, understanding, and improving student engagement in programming education. We investigate whether synthetic data generation can help identify at-risk students earlier in a small, imbalanced dataset from an introductory programming module. The analysis used anonymised records from 379 students, with 15\% marked as failing, and applied several machine learning algorithms. The first experiments showed poor recall for the failing group. However, using synthetic data generation methods led to a significant improvement in performance. Our results suggest that machine learning can help identify at-risk students early in programming courses when combined with synthetic data. This research lays the groundwork for validating and using these models with live student cohorts in the future, to allow for timely and effective interventions that can improve student outcomes. It also includes feature importance analysis to refine formative tasks. Overall, this study contributes to developing practical workflows that help detect disengagement early and improve student success in programming education.

\end{abstract}
\section{Introduction}
This study forms part of a wider project focused on measuring, understanding, and improving student engagement in programming education. An early challenge was the difficulty of predicting disengagement and failure in small, single-module datasets, where failing students are few. We collected data from three cohorts of an introductory programming module at Coventry University, comprising 379 student records, with 85\% passing and 15\% failing. The dataset captures performance across up to 150 formative coding tasks, automatically marked through our learning management system.

While task completion and time-on-task are established engagement indicators, much existing research relies on larger, multi-module datasets or demographic information, often unavailable due to privacy constraints. Instead, our work focuses on coding task performance within a single module. Initial trials with classification algorithms (e.g., logistic regression, random forests, SVMs) showed poor recall for failing students, limiting early intervention potential. To address this, we explore synthetic data generation techniques to rebalance the dataset, aiming to strengthen \emph{at-risk prediction} and lay the foundation for future live validation and interventions. This study addresses the following questions:
\begin{itemize}
    \item \textbf{RQ1:} How can synthetic data enhance the prediction of at-risk students in small, imbalanced datasets?
    \item \textbf{RQ2:} Which machine learning algorithms are most effective for this task?
\end{itemize}

\section{Related Work} 

Studies on computer science education have documented consistently low success rates in introductory programming courses. Pass rates typically hover around 67\%, with larger class sizes and the use of languages such as C associated with increased failure rates \cite{watson_failure_2014}. Research on student engagement identifies behavioural, cognitive, and affective dimensions as relevant to academic performance \cite{kahu_framing_2013,kahu_student_2018}. Metrics such as time-on-task and assessment completion are commonly used to quantify engagement. Institutional reliance on retrospective surveys (e.g., the UK National Student Survey and module evaluations) limits opportunities for real-time prediction \cite{fredricks_school_2004}.

Automation has increased the volume of formative feedback available for programming tasks. At Coventry University, automated feedback was associated with improved satisfaction metrics, though the effects were less pronounced for lower-performing students \cite{croft_computing_2019}. Students also reported consistent value in receiving human feedback \cite{grawemeyer_feedback_2022}. Keuning et al. \cite{keuning_systematic_2019} found that solution-focused error messages support syntactic and conceptual understanding but often do not address strategic knowledge, which remains more effectively conveyed through human instruction.

Machine learning techniques have been applied to engagement data to predict students at risk of failure \cite{qian_students_2017}. Some approaches incorporate demographic variables, though access to such data may be limited by institutional and legal constraints \cite{dervenis_predicting_2022}. Olney et al. \cite{olney_student_2024} developed a predictive framework analysing academic performance and engagement over time to identify students needing intervention. This study focuses on real-time prediction within a single programming module by modelling student performance on formative tasks.

\section{Methodology}

\subsection{Dataset}

An anonymised dataset from three cohorts comprised 379 unique student records. Summative grades were used to assign a Boolean label: passing (\texttt{true}) or failing (\texttt{false}), with an imbalanced distribution of approximately 85:15 in favour of passing. This included all available data for the module.

Formative task submissions were recorded via the Codio learning platform (\url{https://www.codio.com}), which automatically marks tasks. Most involve writing, extending, or debugging small Python programs evaluated against unit tests, though some include other formats (e.g., multiple choice questions). Students are allowed multiple attempts on programming tasks. For further details, see \cite{croft_computing_2019}.

\subsection{Data Preparation}

For each student, we compile two comma-separated lists, \texttt{right\_answers} and \texttt{wrong\_answers}, containing the corresponding task names.  These are then converted into one-hot-encoded (OHE) features.  
We use three different feature sets corresponding to three different engagement intervals: 43 features for Weeks 1--3, 106 features for Weeks 1--6, and 150 features for Weeks 1--9. 
In this encoding, a value of \texttt{1} indicates that a task was completed correctly, while \texttt{0} means that it was either not attempted or not solved correctly (even after multiple attempts). Following this, a train--test split of 80:20 was used to provide the synthetic data generation process with as much data as possible.

\subsection{Baseline Results (Pre-Synthetic Data Generation)}

Baseline experiments evaluated several classification algorithms to establish a reference point for synthetic data approaches. Models included Logistic Regression, SVMs (RBF and linear), Decision Trees, Random Forests, Naive Bayes, and K-Nearest Neighbours, all using default \texttt{sklearn} settings on the original imbalanced dataset. Table~\ref{tab:baseline} reports the best-performing model for the failing (\texttt{false}) class at each interval, with corresponding precision, recall, and F1-scores.

\begin{table}[ht]
\centering
\caption{Baseline Performance for the Failing Class (Pre-Synthetic Data)}
\label{tab:baseline}
\begin{tabular}{lccccc}
\hline
\textbf{Interval} & \textbf{Best Model} & \textbf{Precision (\texttt{false})} & \textbf{Recall (\texttt{false})} & \textbf{F1-Score (\texttt{false})} \\
\hline
1$-$3 & Decision Tree         & 0.50 & 0.45 & 0.48 \\
1$-$6 & Naive Bayes           & 0.47   & 0.64 & 0.54 \\
1$-$9 & Logistic Regression   & 0.50   & 0.36 & 0.42 \\
\hline
\end{tabular}
\end{table}

The highest F1-score for the failing class (\texttt{false}) was 0.5385, achieved by Naive Bayes during Weeks 1$-$6. However, consistently low precision and recall across all intervals suggest these baseline models are inadequate for early intervention. 

\subsection{Synthetic Data Generation}

We applied three synthetic data techniques: SMOTE \cite{chawla_smote_2002}, ADASYN \cite{haibo_he_adasyn_2008} (both oversampling methods using different interpolation strategies), and CTGAN \cite{xu_modeling_2019}, a generative approach. Each method created a balanced 50/50 class distribution without altering feature dimensions.

Classifiers were retrained on these balanced datasets, while the test split containing only real data remained unchanged. Table~\ref{tab:postsynthetic} presents the best-performing model for the failing class at each interval, with corresponding precision, recall, and F1-score.

\begin{table}[ht]
\centering
\caption{Post Synthetic Data Performance for the Failing Class}
\label{tab:postsynthetic}
\begin{tabular}{lcccc}
\hline
\textbf{Interval} & \textbf{Best Model} & \textbf{Precision} & \textbf{Recall} & \textbf{F1-Score} \\
\hline
1$-$3 & ADASYN/SVM (RBF) & 0.60   & 0.55 & 0.57 \\
1$-$6 & SMOTE/LR  & 0.70   & 0.64 & 0.67 \\
1$-$9 & SMOTE/LR  & 0.60   & 0.55 & 0.57 \\
\hline
\end{tabular}
\end{table}

Synthetic data augmentation improved recall and F1-scores for the failing class, with the best performance during Weeks 1$-$6. CTGAN underperformed compared to the oversampling methods across all models and intervals. Not shown in the table, SMOTE/LR in Weeks 1$-$3 achieved a higher recall of 0.6364 and a competitive F1-score of 0.56 for the \texttt{false} label. Therefore, SMOTE/LR was used to further tune across all intervals for consistency.

\subsection{Synthetic Data Visualization}

\begin{figure}[!htb]
\centering
\begin{minipage}[b]{0.3\textwidth}
  \centering
  \includegraphics[width=\textwidth]{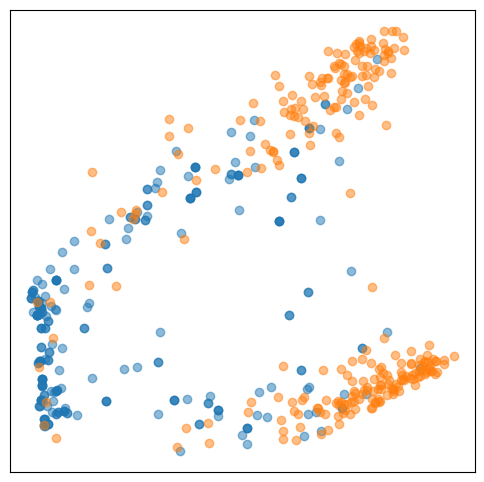}
  \vspace{0.5em}
  \centerline{(a) SMOTE}
\end{minipage}\hfill
\begin{minipage}[b]{0.3\textwidth}
  \centering
  \includegraphics[width=\textwidth]{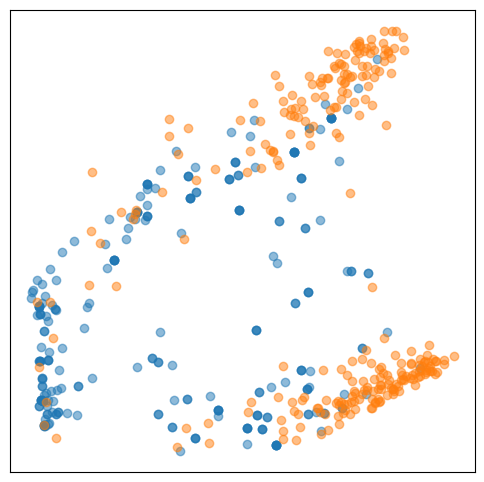}
  \vspace{0.5em}
  \centerline{(b) ADASYN}
\end{minipage}\hfill
\begin{minipage}[b]{0.3\textwidth}
  \centering
  \includegraphics[width=\textwidth]{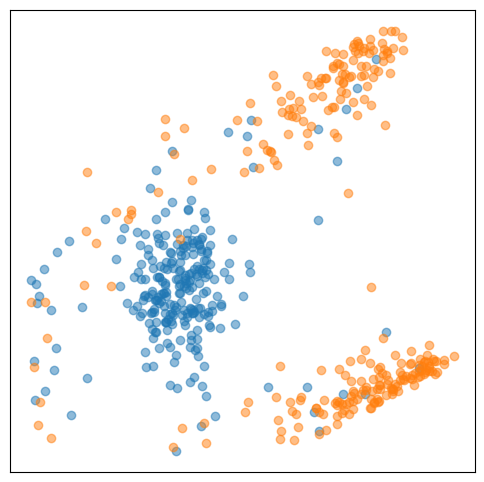}
  \vspace{0.5em}
  \centerline{(c) CTGAN}
\end{minipage}
\caption{PCA visualisations of synthetic samples generated. Blue (\texttt{false}), Orange (\texttt{true})}
\label{fig:synthetic_pca}
\end{figure}

To investigate CTGAN’s poor performance, we applied Principal Component Analysis (PCA) to visualise the high-dimensional feature space (Figure~\ref{fig:synthetic_pca}) using data from Weeks 1$-$6, representative of all intervals. SMOTE and ADASYN generated synthetic samples that aligned well with original data clusters, preserving class structure. In contrast, CTGAN showed mode collapse, with samples clustering tightly and failing to reflect the minority class’s variability.

\section{Results}
We tuned SMOTE/LR models using grid search to maximise performance, with results shown in Table~\ref{tab:tuned_metrics}.

For Weeks 1--3, the best setup used SMOTE (\texttt{k\_neighbors=5}, sampling strategy `auto') with elasticnet Logistic Regression (\texttt{penalty=`elasticnet', l1\_ratio=0.5, C=0.01, solver=`saga'}) at a threshold of 0.50, yielding precision, recall, and F1-score of 0.55. This improved precision over baseline (0.50) but slightly reduced recall (-0.05).

For Weeks 1$-$6, the optimal model used SMOTE (\texttt{k\_neighbors=3}) with Logistic Regression (\texttt{penalty=`l2', C=1.0, solver=`lbfgs'}) at a threshold of 0.45, reproducing baseline precision of 0.70 and recall of 0.64.

Weeks 1$-$9 used a setup similar to Weeks 1$-$3: SMOTE (\texttt{k\_neighbors=5}) and elasticnet Logistic Regression (\texttt{C=1.0, l1\_ratio=0.5, solver=`saga'}) at a threshold of 0.50, achieving 0.75 precision, 0.55 recall, and 0.64 F1-score—marking a notable precision gain over the baseline (0.60) with stable recall and F1.

\begin{table}[ht]
\centering
\caption{Tuned Metrics and Best Parameters for LR--SMOTE }
\label{tab:tuned_metrics}
\begin{tabular}{lccccc}
\hline
\textbf{Interval} & \textbf{Precision} (\texttt{false})  & \textbf{Recall} (\texttt{false}) & \textbf{F1 Score} (\texttt{false})& \textbf{Accuracy} & \textbf{AUC} \\
\hline
1--3 & 0.55 & 0.55 & 0.55 & 0.87 & 0.77 \\
1--6 & 0.70 & 0.64 & 0.67 & 0.91 & 0.78 \\
1--9 & 0.75 & 0.55 & 0.64 & 0.91 & 0.74 \\
\hline
\end{tabular}
\end{table}

\section{Conclusion and Future Work}
This study explored whether synthetic data generation can improve the identification of at-risk students in a small, imbalanced dataset from a single introductory programming module. Oversampling techniques such as SMOTE and ADASYN improved recall and F1 scores for the failing class compared to baseline models, addressing \textbf{RQ1}. Preliminary analysis also indicated that performance on specific formative tasks may predict failure, supporting future feature importance analysis for \textbf{RQ2}.

The next phase will validate the models using live cohort data currently being collected. If results hold, the models will be deployed for real-time prediction and early intervention over the following two modules. Similarly, feature importance analysis will help refine formative tasks and guide instructional improvements. While this study focused on a single module, the methodology will be extended to additional programming modules delivered through Codio to explore generalisability. The approach may apply to any context using numerous automatically assessed micro-tasks, particularly where richer datasets are unavailable.

This work forms an essential step toward developing practical, interpretable workflows to help educators detect disengagement early and support student success in programming education.


\bibliographystyle{splncs04}
\bibliography{references}

\end{document}